# BiLingual Information Retrieval System for English and Tamil

Dr.S.Saraswathi, Asma Siddhiqaa.M, Kalaimagal.K, Kalaiyarasi.M

**Abstract-**This paper addresses the design and implementation of BiLingual Information Retrieval system on the domain, Festivals. A generic platform is built for BiLingual Information retrieval which can be extended to any foreign or Indian language working with the same efficiency. Search for the solution of the query is not done in a specific predefined set of standard languages but is chosen dynamically on processing the user's query. This paper deals with Indian language Tamil apart from English. The task is to retrieve the solution for the user given query in the same language as that of the query. In this process, a Ontological tree is built for the domain in such a way that there are entries in the above listed two languages in every node of the tree. A Part-Of-Speech (POS) Tagger is used to determine the keywords from the given query. Based on the context, the keywords are translated to appropriate languages using the Ontological tree. A search is performed and documents are retrieved based on the keywords. With the use of the Ontological tree, Information Extraction is done. Finally, the solution for the query is translated back to the query language (if necessary) and produced to the user.

———————— ◆ ————————

## 1. Introduction

**Cross-language information retrieval (CLIR)** is a subfield of information retrieval dealing with retrieving information written in a language different from the language of the user's query.

The development of information repositories is creating many opportunities and also new challenges in information retrieval. The availability of online documents in many different languages makes it possible for users around the world to directly access previously unimagined sources of information. However in conventional information retrieval systems the user must enter a search query in the language of the documents in order to retrieve it. This requires that users can express their queries in those languages in which the information is available and can understand the documents returned by the retrieval process. This restriction clearly limits the amount and type of information that an individual user really has access to.

*Cross-Language Information Retrieval* [1] lets users express their query in their native language, and irrespective of the language in which the information is available, present the information in the user-preferred language or set of languages, in a manner that satisfies the user's information needs.

*Cross Language Evaluation Forum (CLEF)* aims at promoting research in the design of multi- lingual, multi-modal retrieval systems by providing an opportunity for the research communities working in different languages to collaborate and share their experiences.

All the existing systems are highly language dependant. Extension of such a system to include a new language either tends to be unfeasible ([2] works only for cognate languages-Hindi and Punjabi) or feasible with a minimum of two translations requiring ([3] makes use of bilingual dictionary).

Most of the existing systems perform a search for the solution in a standard set of languages. For example, [4] searches for the solution only in English documents. This does not turn out to be successful in all cases because the domain of the query may be related to a particular region where a particular language is spoken.

So searching in that regional language will undoubtedly lead to a more precise solution. The proposed system implements this concept.

## 2 Modules of Proposed System

### 2.1 Pre Requisites

Usage of the Ontological tree is the striking feature of this system. An Ontological tree can be defined as entities and grouping of entities, classification of entities

———————————————

The authors acknowledge All India Council of Technical education (AICTE) which has funded under Research Promotion Scheme for carrying out this project.

- *F.A. Author is Associate Professor, in the Department of Information Technology, Pondicherry Engineering College, Pondicherry, India.*

- *S.B. Author are Under graduate students in the Department of Information Technology, Pondicherry Engineering College, Pondicherry, India.*

in a hierarchical order and grouping of entities based on similarities and differences.

Ontological trees [5] are used in the phase of dynamically deciding the search language for the given query and retrieving the related keywords in the language chosen for document search.

For this process, building of an ontological tree is required satisfying a number of constraints as listed below:
- Nodes with multiple entries
- Entry for both the languages in every node

This tree is also used to find the language which is most appropriate to the content of the user given query. To do so, in every node a separate attribute is set, to give information about the root language of the word present in that node.

As a prototype of a search done on an online search engine, information extraction is done from a set of collected documents on the specified domain. Hence collection of documents pertaining to the domain is another pre requirement in this process.

**2.2 System Development**

The entire process can be divided into 4 main modules. They are **User Interface, Keywords Extraction, Information Retrieval and Information Extraction and Output display.**

The detailed description of the modules is given below.

**Module 1: User Interface**

This is the first stage of implementation. The main objective of this module is to get a query from the user whose solution is to be found.

The various sub modules under this are as follows

**Language selection :** The user is initially asked to choose one among the 2 languages namely : Tamil or English.

The user can select a language in which he is most comfortable. The output of the query will also be in the language chosen by the user.

**User's query :** After the user selects a language, a virtual keyboard appropriate to the selected language is displayed on the screen . By typing on the virtual keyboard the user can enter the query in his selected language.

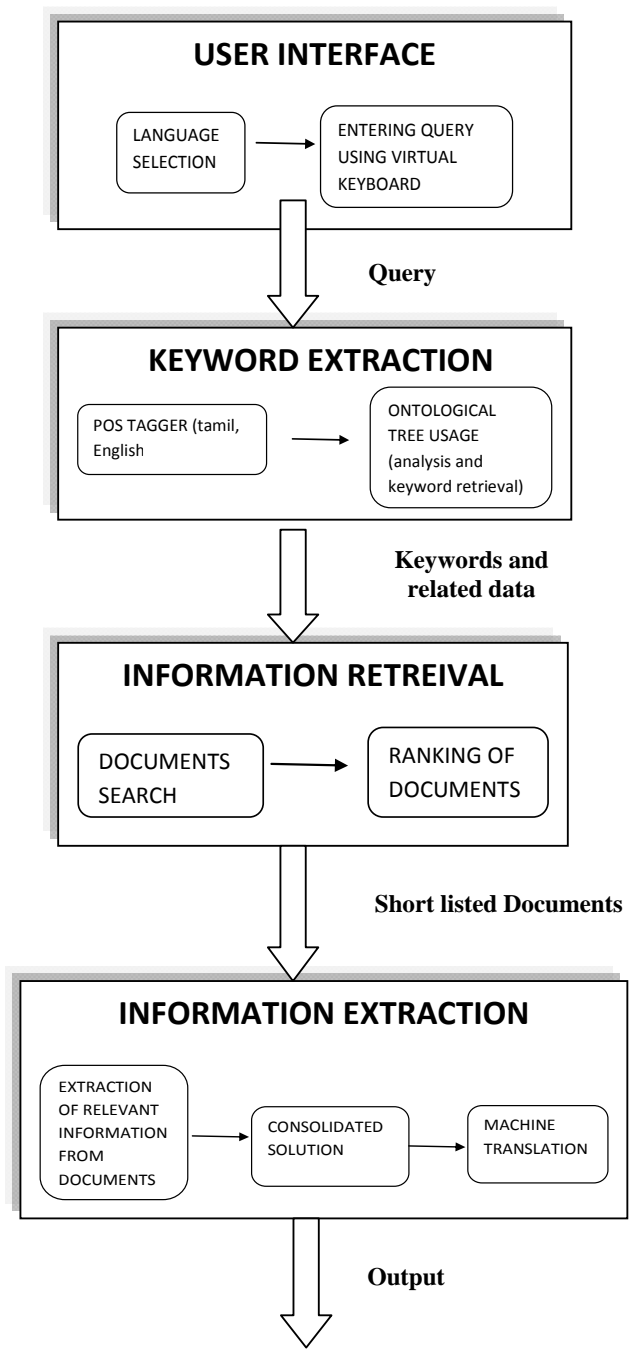

Fig 1: Module Flow Diagram

**Module 2 : Keyword Extraction**

The input to this module is the user given query and the output is the keywords and the related data that are obtained from the input query.

**Tagger Usage:** A Tagger is software that takes a sentence as input, separates the words in that sentence, tries to identify the parts of the sentences like nouns, verb, preposition etc. and returns every word in the sentence along with its type as an output.

The Query given by the user is fed as an input to the appropriate Tagger. It tags the input sentence and returns the parts of the sentences. From the output of the tagger, verbs and nouns are identified and are set as main keywords to perform the document search.

**Usage of Ontological Tree:** Each keyword identified is matched with entry in every node in the Ontological tree. The exact location if the keyword in the tree is identified.
The language which is most appropriate to the keyword is traced with the help of the attribute set in that node and the language of the document search is identified.

All the child nodes of the keyword node are traversed and their corresponding entries in the above decided language are noted as related data for document search.

**Module 3 : Information Retrieval**

Information retrieval is the process of retrieving relevant information from a pool of data. The input to this module is the keywords obtained from the ontological tree and the output of this module is a retrieved set of short listed documents.

The sub modules here are as follows:

**Document Search:** A set of documents have already been collected in the pre requisite processing section. Based on the occurrences of the keywords in each document, a set of documents are short listed.

The algorithm used for this purpose is the Naïve Algorithm. The naive algorithm is used to search every location of the string *t* for the pattern *p*. In this case, the document is considered as *t*, the keywords as *p* and *t* is searched for *p*.

**Ranking of Documents:** The documents that are retrieved are ranked based on the relevance of the search words on the document. For this purpose, the PageRank algorithm is used. PageRank is an algorithm that assigns a numerical weighting to each document with the purpose of measuring its relative importance in the domain.

**Module 4 : Information Extraction**

The Input to this module is the set of short listed documents and the output is the solution for the user given query which is displayed finally to the user.

**Extraction of Relevant Information From Documents:** To locate the apt solution for the query, search is done in the short listed documents. This is done on all the short listed documents and the exact solution for the user's query is found.

**Consolidated Solution:** Now a set of solutions extracted from the shorted listed documents are available. All the possibilities are now consolidated, avoiding repetition and apt solutions are retrieved.

**Machine Translation:** Machine Translation (MT) investigates the use of computer software to translate text or speech from one natural language to another.

The language in which the solution is to be retrieved is checked with the user chosen language. If found to be the same, the solution is directly presented to the user.

Else, Machine Translation is done with the use of the ontological trees, to re convert the solution that is found back to the user's query language and then presented to the user.

**2.3 Parameters For Testing:**

The working of the software will be tested by considering the following parameters.

**Precision :** Precision is defined as the number of relevant documents retrieved by a search divided by the total number of documents retrieved by that search

$$\text{precision} = \frac{|\{\text{relevant documents}\} \cap \{\text{retrieved documents}\}|}{|\{\text{retrieved documents}\}|}$$

(1)

Precision takes all retrieved documents into account, but it can also be evaluated at a given cut-off rank, considering only the topmost results.

**Recall :** Recall is defined as the number of relevant documents retrieved by a search divided by the total number of existing relevant documents (which should have been retrieved).

$$\text{recall} = \frac{|\{\text{relevant documents}\} \cap \{\text{retrieved documents}\}|}{|\{\text{relevant documents}\}|}$$

(2)

Recall in Information Retrieval is the fraction of the documents that are relevant to the query that are successfully retrieved. It can be looked as the probability that a relevant document is retrieved by the query.

**F-measure :** A measure that combines Precision and Recall is the harmonic mean of precision and recall, the traditional F-measure or balanced F-score:

$$F = 2 \cdot \frac{\text{precision} \cdot \text{recall}}{\text{precision} + \text{recall}}. \quad (3)$$

It considers both the precision $p$ and the recall $r$ of the test to compute the score: $p$ is the number of correct results divided by the number of all returned results and $r$ is the number of correct results divided by the number of results that should have been returned.

## 3. Comparison With Existing Systems

### 3.1 Ontology Versus Dictionary:

Nearly all of the systems that exist have made use of the Dictionary (mostly a Bilingual Dictionary) for the purpose of translation. This makes the system highly language dependant. Hence extension of the system reduces the efficiency of the system.

In turn, Ontological tree has inclusions of any number of languages with no restriction. Also, only a single mapping will be required from any language to any other language.

Thus the Ontological tree built is used for a number of other purposes also such as keyword language identification and sub keywords extraction. These cannot be fulfilled by a dictionary.

### 3.2 Restriction Of Search Languages:

All of the existing Cross Lingual Information Retrieval Systems concentrate on making the required information available in a desired language and not exactly on obtaining the perfectly apt solution for the given query in an appropriate language. All such systems existing so far search for the solution only in documents same as that of the query language or in a predefined set of languages. They do not search in the languages which are more relevant to the given query.

### 3.3 Complexity Of The System:

Few systems have used the concept of ontology to retrieve the keywords which are relevant to the main keywords under search, but all of them use ontological trees with a single entry in every node which leads to the usage of a voluminous tree increasing the space complexity of the system considerably. No ontological tree having multiple entries at every node exists so far.

They have used different techniques and different concepts at various phases of the query solution retrieval. Implementation of all these, tend to increase the space complexity of the system as the whole. The proposed design uses just the single tree for all the purposes.

## 4. Results

The solution obtained for both Tamil and English under the domain "Festival" are given below. Documents related to Christmas, New Year, Easter and Good Friday were collected for English language and documents related to Diwali, Pongal, Navarathiri, Kaarthigai Deepam and Vinayagar chathurthi were collected for Tamil language. Total of 200 documents were collected for both the languages. The following are the results for monolingual English and Tamil and Bilingual English and Tamil.

Table 1 : Analysis of results for monolingual English

| QUERY TYPE | REVELANT DOCUMENTS | PRECISION | RECALL | F-MEASURE |
|---|---|---|---|---|
| what is the actual birth date of jesus | 58 | 0.631 | 1.000 | 0.773 |
| what did charles dickens say about Christmas | 46 | 0.556 | 0.761 | 0.642 |
| what does new testament say about easter | 58 | 0.591 | 0.672 | 0.629 |
| when was the Crucifixion of Jesus | 58 | 0.602 | 0.918 | 0.727 |
| when do romans celebrate new year | 61 | 0.711 | 0.967 | 0.819 |

Table 2 : Analysis of results for monolingual Tamil

| QUERY TYPE | REVELANT DOCUMENTS | PRECISION | RECALL | F-MEASURE |
|---|---|---|---|---|
| காணும் பொங்கல் எவ்வாறு கொண்டாடப்படுகிறது | 6 | 0.431 | 1.000 | 0.602 |
| தீபாவளி எதனால் கொண்டாடப்படுகிறது | 21 | 0.408 | 0.952 | 0.571 |
| நவராத்திரி பற்றி விவரி | 18 | 0.295 | 1.000 | 0.456 |
| கார்த்திகை தீபம் எவ்வாறு கொண்டாடுவர் | 13 | 0.371 | 1.000 | 0.541 |
| பொங்கல் எப்பொழுது | 22 | 0.550 | 1.000 | 0.710 |

Table 3 : Analysis of results for bilingual English-Tamil

| QUERY TYPE | APT LANGUAGE | REVELANT DOCUMENTS | PRECISION | RECALL | F-MEASURE |
|---|---|---|---|---|---|
| explain about crackers | T | 9 | 1.000 | 0.889 | 0.941 |
| Different day of pongal | T | 22 | 0.564 | 1.000 | 0.721 |
| கிறிஸ்துமஸ் பரிசுகள் | E | 46 | 0.571 | 0.783 | 0.661 |
| ஏசுநாதர் சிலுவை | E | 56 | 0.625 | 0.948 | 0.753 |

The System was tested for information retrieved using the keywords alone and with the use of concept words obtained from the Ontology. It was found that the relevance was improved by 40% for English and 60% for Tamil language as shown in Figure 2.

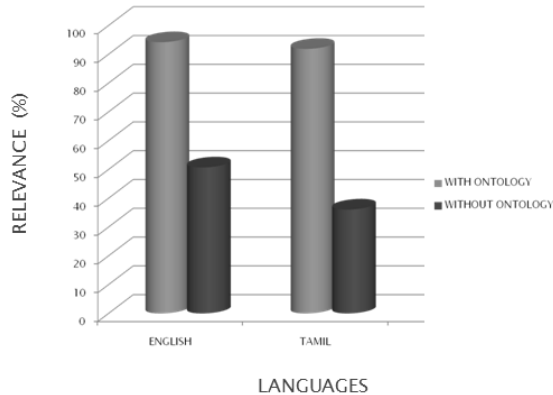

Figure 2: Comparison of results with and without Ontology

## 5. Conclusion

BiLingual Information Retrieval Systems play a very vital role in a country like India having a wide variety of regional languages. Extension of such systems to either a foreign language or an Indian language is also a must. Thus the system whose design is proposed aims at obtaining the solution through search done in dynamically chosen apt languages, uses Ontological tree for multiple purposes like inter language conversion, keyword language identification and sub keywords extraction. Also, a generic platform for incorporating any other language is also obtained.

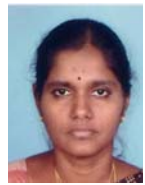

First A. Author S.Saraswathi is Assistant professor, in the Department of Information Technology, Pondicherry Engineering College, Pondicherry, India. She completed her PhD, in the area of speech recognition for Tamil language. Her areas of interest include speech processing, artificial intelligence and expert systems.

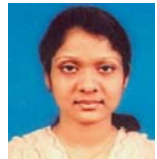

**Second B. Author Jr Asma Siddhiqaa M is** final year student of Department of Information Technology, Pondicherry Engineering College, Pondicherry, India. Her areas of interest are Artificial Intelligence, Robotics

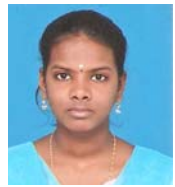

**Second B. Author Jr Kalaimagal K** is final year student of Department of Information Technology, Pondicherry Engineering College, Pondicherry, India. Her areas of interest are Artificial Intelligence, Object oriented Programming concepts

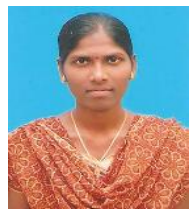

**Second B. Author Jr Kalaiyarasi M** is final year student of Department of Information Technology, Pondicherry Engineering College, Pondicherry, India. Her areas of interest are Artificial Intelligence, Data Structures.